# BLACK HOLE ENTROPY CONSTRAINTS ON VARIATION OF THE GRAVITATIONAL CONSTANT

JANE H. MACGIBBON
Dept of Physics and Chemistry
University of North Florida
Jacksonville FL 32224

June 15 2007

## ABSTRACT

**Here we apply the Generalized Second Law of Thermodynamics (GSL) to black holes accreting and emitting in the present Universe and derive upper limits on the variation in the gravitational constant G. The limits depend on how the gravitational mass M varies with G. Parameterizing $M \propto G^n$, if $n > -1/2$ (including $n = 0$), the GSL applied to the full range of black holes theoretically allowed in the present Universe does not constrain an increase in G but any decrease must be less than $|G^{-1} dG/dt| \approx 10^{-52}$ per second. If $n < -1/2$, the GSL does not constrain a decrease in G but any increase must be less than $|G^{-1} dG/dt| \approx 10^{-52}$ per second. At earlier redshifts, these constraints weaken as $z^3$. If $n = -1/2$, the GSL does not constrain a decrease but any increase must be less than $|G^{-1} dG/dt| \approx t^{-1}$. If the mass range is restricted to those black holes which have been astronomically observed, the present constraints on $n > -1/2$ and $n < -1/2$ are only weakened by a factor of $\sim 10^8$ with the tightest constraints coming from stellar mass black holes and the $n = -1/2$ bound does not change. The stellar mass black hole limits should constrain the variation of G in Standard Model physics and all extension models which approximate classical physics on astronomical scales.**

Numerous recent experimental and theoretical work has raised the possibility that fundamental constants may be changing as the Universe ages. By applying the Generalized Second Law of Thermodynamics (GSL) to black holes accreting and emitting in the present Universe, we can derive upper limits on the variation in the constants. In the previous paper in this series[1], we extended the proposal of P.C.W. Davies et al. [2] by including the full description of the time variation of the entropy of the black hole system and derived an upper limit on the variation in the electronic charge $e$ of $de/dt \approx 10^{-23} e$ per second by considering the full range of theoretically allowed charged black holes in the present Universe. In this paper, we present our second result using the GSL methodology: the limits on the variation permitted in the gravitational constant $G$. We will present two sets of limits on $G$: one derived using the full range of theoretically allowed black holes in the present Universe and the slightly weaker set of limits derived using only the mass range of black holes so far observed by astronomers. Throughout this paper we assume that the speed of light $c$, Planck's constant $\hbar$ and Boltzmann's constant $k$ are constant and investigate variation in $G$. Extension of this methodology to dependent or independent variation in the other fundamental constants is straightforward and will be presented elsewhere.

The Generalized Second Law of Thermodynamics, derived for black hole systems, states that the net entropy of the system can not decrease with time [3]. Over a time interval $\Delta t$, the net generalized entropy of the system increases by

$$\Delta S = \Delta S_{BH} + \Delta S_{R+M} \geq 0 \qquad (1)$$

where $\Delta S_{BH}$ and $\Delta S_{R+M}$ are the change in entropy of the black hole and of the ambient radiation and matter, respectively. The entropy of a black hole is

$$S_{BH} = \frac{kc^3}{4\hbar G} A_{BH} \qquad (2)$$

where $k$ is the Boltzmann constant. For an uncharged, non-rotating (Schwarzschild) black hole of mass $M$, the area of the black hole is

$$A_{BH} = \frac{16\pi G^2 M^2}{c^4}. \qquad (3)$$

Hawking [4,5] has established that a black hole is continuously emitting quasi-thermal radiation with a temperature

$$T_{BH} = \frac{2\hbar GM}{kcA_{BH}} = \frac{\hbar c^3}{8\pi kGM}. \qquad (4)$$

Thus $\Delta S_{BH}$, the full change in black hole entropy over time $\Delta t$, must include the contribution from the Hawking flux as well as any partial change induced by a variation in $G$, i.e.

$$\Delta S_{BH} \approx \frac{dS_{BH}}{dt}\Delta t = \left\{\left(\frac{\partial S_{BH}}{\partial G} + \frac{\partial S_{BH}}{\partial M}\frac{\partial M}{\partial G}\right)\frac{dG}{dt} + \frac{\partial S_{BH}}{\partial M}\frac{dM_H}{dt}\right\}\Delta t. \qquad (5)$$

where the subscript $H$ denotes Hawking radiation (or where appropriate thermal accretion). The black hole mass will decrease as the black hole radiates.

Immediately we are presented with the question: does the gravitational mass of an object $M$ depend on $G$? To take account of the possibility that $M$ may depend on $G$, let us write $M \propto G^n$ where $n$ may be 0, negative or positive. Then $\partial M / \partial G = nM / G$ and

$$\frac{dS_{BH}}{dt} = \frac{8\pi kGM}{\hbar c}\left\{\left(n+\frac{1}{2}\right)\frac{M}{G}\frac{dG}{dt} + \frac{dM_H}{dt}\right\} \qquad (6)$$

To proceed further we must consider the two cases when the black hole temperature is greater than and less than the temperature of its surroundings.

Case (I) If the black hole temperature is greater than the temperature of its surroundings, ie $T_{BH} > T_{R+M}$, there will be a net radiation loss from the black hole into its environment. The mass loss due to Hawking radiation is [6]

$$\frac{dM_H}{dt} \approx -\frac{\hbar c^4}{G^2 M^2}\beta \qquad (7)$$

with $\beta \approx 3\text{x}10^{-4}$ for a hole emitting the photon, 3 light neutrino species and the graviton. D.N. Page [7] has numerically calculated that for a black hole emitting the photon, 3 light neutrino species and the graviton the increase in $S_{R+M}$ due to Hawking emission is 1.62 times the corresponding decrease in $S_{BH}$ due to Hawking emission. Thus $\Delta S \geq 0$ provided the first term within the {} brackets in Eq (6) is not negative and of order the second term.

Applying this constraint, the limit we seek breaks into 5 subcases: (i) if $n > -1/2$ then there is no constraint on the rate at which $G$ may increase ie on $dG/dt > 0$; (ii) if $n < -1/2$ there is no constraint on the rate at which $G$ may decrease ie on $dG/dt < 0$. In the case of either (iii) $n < -1/2$ and $G$ increasing or (iv) $n > -1/2$ and $G$ decreasing, we require

$$\left|\left(n+\frac{1}{2}\right)\frac{dG}{dt}\right| \lesssim \frac{0.6\hbar c^4 \beta}{GM^3} \tag{8}$$

Any change to the emission rate Eq (7) due to varying $G$ is a small higher-order term and will not affect this bound. The tightest constraint in Eq (8) is set by the largest allowed black hole mass such that the black hole is still Hawking radiating into its environment. For the present Universe, this mass is $M_{CMB} \approx 4.5 \times 10^{25}$ g, the mass of a black hole whose temperature equals the 2.73K cosmic microwave background. Thus in subcases (iii) and (iv), the entropy constraint corresponds to

$$\left|\left(n+\frac{1}{2}\right)G^{-1}\frac{dG}{dt}\right| \lesssim \frac{0.6\hbar c^4 \beta}{G^2 M_{CMB}^3} = \frac{0.6 m_{pl}^3 \beta}{t_{pl} M_{CMB}^3} \approx 4 \times 10^{-52} \text{ s}^{-1} \tag{9}$$

As the Universe ages, $M_{CMB}$ increases and so the magnitude of the maximum allowed rate of variation of $G$ decreases. Conversely, looking back in time to the earlier Universe the bound on $|G^{-1}dG/dt|$ weakens and grows with redshift as $z^3$.

In subcase (v), $n = -1/2$ and the $dG/dt$ term in Eq (15) vanishes. To find a limit for $n = -1/2$ we must include the next term in the Taylor expansion of $\Delta S \approx \Delta t \, dS/dt + (\Delta t)^2 d^2S/dt^2/2 + \ldots$. Proceeding using $dS_{BH}/dt$ as given in Eq (5) and noting that $dM_H/dt$ changes very slowly with respect to $t$, we find for $n = -1/2$ there is no constraint on $G$ decreasing but any increase is bounded by $|G^{-1}dG/dt| \lesssim 4t^{-1}$.

Case (II) If the black hole temperature is less than or equal to the temperature of its surroundings, i.e. $T_{BH} \leq T_{R+M}$, the black hole will accrete from its surroundings faster than it Hawking radiates. This accretion increases the black hole mass $M$, further lowering $T_{BH}$, and leads in turn to more accretion. (As Hawking has pointed out [3], a black hole can not be in stable thermal equilibrium if an unbounded amount of energy is available in its surroundings.) The general thermodynamical definitions of the temperature of the environment and the black hole temperature are respectively [3]

$$T_{R+M}^{-1} \equiv \frac{\partial S_{R+M}}{\partial E}, \qquad T_{BH}^{-1} \equiv c^{-2}\left(\frac{\partial S_{BH}}{\partial M}\right) \tag{10}$$

where $E$ is the energy of the environment. During accretion, the black hole mass increases by an amount equal to the decrease in $E$. Hence for $T_{BH} \leq T_{R+M}$, the temperature definitions imply that the increase in black hole entropy due to accretion must be greater than the decrease in $S_{R+M}$ due to accretion. Also for $T_{BH} \leq T_{R+M}$, the increase in $S_{BH}$ due to accretion must be greater than the decrease in $S_{BH}$ due to Hawking radiation. In analogy with a classical blackbody, a cold large black hole in a warm thermal bath will absorb energy at a rate $dE/dt = \pi^2 k^4 \sigma_S T_{R+M}^4 / 60\hbar^3 c^2$ (and emit radiation $dE/dt = \pi^2 k^4 \sigma_S T_{BH}^4 / 60\hbar^3 c^2$) per polarization or helicity eigenstate where $\sigma_S = 27\pi G^2 M^2 / c^4$ is the geometrical optics cross-section [6]. (Since the entropy of the background is maximized for a thermal bath, a thermal bath will give the strictest accretion constraint on $\Delta S$.) For accretion, Eq (7) is thus replaced by

$$\frac{dM}{dt} = +\frac{\beta_{R+M} \hbar c^4}{G^2 M_{R+M}^2} \left(\frac{M}{M_{R+M}}\right)^2 \tag{11}$$

where $\beta_{R+M} \sim 10^{-4}$ and $M_{R+M}$ is the mass of a black hole whose temperature equals the ambient temperature. Again $\Delta S \geq 0$ provided the first term within the {} brackets in Eq (6) is not negative and of order the second (now absorption) term. Hence the limit breaks into the five subcases: (i) if $n > -1/2$ there is no constraint on the rate at which $G$ may increase; (ii) if $n < -1/2$ there is no constraint on the rate at which $G$ may decrease. In the case of either (iii) $n < -1/2$ and $G$ increasing or (iv) $n > -1/2$ and $G$ decreasing, the magnitude of the rate of change must be less than

$$\left|\left(n+\frac{1}{2}\right) G^{-1} \frac{dG}{dt}\right| \lesssim \frac{\hbar c^4 \beta_{R+M}}{G^2 M_{R+M}^3} \left(\frac{M}{M_{R+M}}\right) \tag{12}$$

Because in the present Universe, $T_{R+M} \gtrsim T_{CMB}$ and so $M_{R+M} \lesssim M_{CMB}$, the tightest constraints are found by considering $M_{R+M} \approx M_{CMB}$ in Eq (12). That is

$$\left|\left(n+\frac{1}{2}\right) G^{-1} \frac{dG}{dt}\right| \lesssim \frac{\hbar c^4 \beta_{R+M}}{G^2 M_{CMB}^3} \left(\frac{M}{M_{CMB}}\right) \tag{13}$$

Also note that for the black hole to (net) accrete from its environment, it must have mass $M \geq M_{CMB}$. Hence, if we consider the full range of theoretically allowed black hole masses in the present Universe, the tightest constraints come from setting $M \approx M_{CMB}$ in Eq (13). The resulting constraints on $|G^{-1} dG/dt|$ are essentially the same as those found in Case (I) for the emitting black holes. Again the bounds weaken as $z^3$ at earlier times. If (v) $n = -1/2$ then considering the next terms in the Taylor expansion for $\Delta S$ we find there is no accretion bound on decreasing $G$ but an increase in $G$ is bounded by $|G^{-1} dG/dt| \lesssim t^{-1}$.

If we restrict the mass range to only those black holes which have been astronomically observed, the tightest constraint of Eq (13) comes from the lightest observed black holes. These are the stellar mass black holes with $M_{stellar} \approx 6 M_\odot \approx 10^{34}$ g for which the bounds for cases (iii) and (iv) become

$$\left|\left(n+\frac{1}{2}\right) G^{-1} \frac{dG}{dt}\right| \lesssim 10^{-44} \text{ s}^{-1} \tag{14}$$

These constraints are only about $10^8$ weaker than the constraints found using $M \approx M_{CMB}$.

Combining Cases (I) and (II), we conclude that the GSL gives the following limits on the variation in the gravitational constant: (i) if $n > -1/2$ then there is no constraint on the rate at which $G$ may increase; (ii) if $n < -1/2$ there is no constraint on the rate at which $G$ may decrease. In the case of either (iii) $n < -1/2$ and $G$ increasing or (iv) $n > -1/2$ and $G$ decreasing, the GSL requires that $\left|G^{-1}dG/dt\right| \lesssim 10^{-52}$ per second; and (v) if $n = -1/2$ there is no constraint on the rate at which $G$ may decrease and $G$ may increase at a rate of $\left|G^{-1}dG/dt\right| \lesssim t^{-1}$. The limits for cases (iii) and (iv) are weaker by about a factor of $10^8$ if we consider only astronomically observed black holes. The constraints on the variation in $G$ in cases (iii) and (iv) are far tighter than all current experimental limits.

Extending our analysis to charged, rotating (Kerr-Newman) black holes does not modify our conclusions. For a rotating, charged (Kerr-Newman) black hole with charge $Q$ and angular momentum $J$,
$A_{BH} = 4\pi G^2 \left(M + \sqrt{M^2 - Q^2/G - c^2 J^2/G^2 M^2}\right)^2 / c^4$ and it is straightforward to show that

$$\frac{\partial S_{BH}}{\partial G} + \frac{\partial S_{BH}}{\partial M}\frac{\partial M}{\partial G} = \frac{2\left(n + \frac{1}{2}\right)\pi k}{\hbar c} \left\{1 + \left(M / \sqrt{M^2 - \frac{Q^2}{G} - \frac{c^2 J^2}{G^2 M^2}}\right)\right\} \times \\ \left(M\sqrt{M^2 - \frac{Q^2}{G} - \frac{c^2 J^2}{G^2 M^2}} + M^2 + \frac{c^2 J^2}{G^2 M^2}\right) \quad (15)$$

Because of the $(n+1/2)$ factor in Eq (15), the GSL applied to rotating charged black holes does not constrain our previously unconstrained cases (i) and (ii). Also it is clear from the form of Eq (15) that the tightest constraints on $\left|G^{-1}dG/dt\right|$ in cases (iii) and (iv) come from considering neutral non-rotating Schwarzschild black holes.

The definitions of $S_{BH}$ and the GSL used above are required for classical General Relativity to be consistent classical Thermodynamics [3,8,9]. Our constraints however should also apply to any extension models which approximate classical physics on the scales relevant to this calculation (which are primarily stellar mass scales). Any modification arising from those extension theories should appear as smaller higher-order corrections in Eq (5) with little change to our conclusions. Note that our approach is also applicable if the variation is in the 'effective' value of the constants, rather than in the underlying 'fundamental' constants. Such a variation could arise, for example, from a higher-order coupling whose strength varies as the Universe ages due to the decreasing cosmological density or temperature. Thus detection of variation need not imply that the Standard Model must be superceded by an extended model which includes variation in the underlying 'fundamental' constants. Because we have embedded the black holes in the ambient Universe, variation in the 'effective' constant should also be constrained by the above limits.

In our entropy calculation we have assumed the standard cosmological picture that the Universe (or more precisely the CMB) is expanding isentropically. Thus $\Delta S_{R+M}$ here contains only the contribution due to black hole accretion and emission and there is no $\Delta S$ contribution from entropic change in the cosmic microwave background as the Universe ages. Implicit in this assumption too is that the isentropic cosmic expansion also compensates for any entropic changes in the cosmic background which would be caused by a varying $G$. (This compensation should also be borne in mind when constructing independent limits on the variation of $G$ from the cosmic microwave background observations.) This implicit assumption should be appropriate because the main (negative) contribution in Eq (6) comes from the change of the intrinsic entropy of a non-relativistic object, the black hole, and not the change in its radiation rate.

A brief discussion of what the value of $n$ may be is also appropriate. Firstly note that in the above it is not required that $M \propto G^n$ over all masses, simply that we can approximate the $G$–dependence of $M$ by $M \propto G^n$ near the mass which gives the tightest constraint. In seeking a value for $n$, the Einstein Equivalence Principle equating inertial mass to gravitational mass requires that $n = 0$, unless the inertial mass also depends on $G$ (as it could, for example, in some Theory of Everything theories which unify gravity and the quantum forces). As we have seen for $n = 0$, the GSL does not limit an increase in G but, of great cosmological interest, any decrease must be less than $\left| G^{-1} dG/dt \right| \approx 10^{-52}$ per second with the bound weakening as $z^3$ at earlier times. If the fundamental constants, or the strength of the 'effective' coupling constants, vary at the maximal rate allowed by the GSL (as proposed in [1]) this variation in G may play an important role in the very Early Universe.

If $n = -1/2$ the GSL bounds on $G$ varying are much weaker, with any increase constrained by $\left| G^{-1} dG/dt \right| \lesssim t^{-1}$. Note however that if $M \propto G^{-1/2}$, the gravitational force between two masses separated by a distance $r_{12}$, $F_G = GM_1 M_2 / r_{12}^2$, remains constant. Because our $n = -1/2$ bounds are found using second-order terms which may be more uncertain, particularly if other higher-order effects must be included, it would be desirable to find a second approach to limit the variation in $G$ in the $n = -1/2$ case, perhaps invoking the inertial mass and $F = ma$. Conversely, experimental searches for variation in $G$ could be used as a test of the GSL and/or of $n$ and/or of the Equivalence Principle and/or of isentropic cosmic expansion (and/or the dependence of variation in $G$ on variation in other constants).

Acknowledgements
It is a pleasure to thank the University of Cambridge for hospitality.[1] J.H. MacGibbon, "Black Hole Constraints on Varying Fundamental Constants" arXiv:0706.2188, accepted by *Phys.Rev.Lett.*
[2] P.C.W. Davies, T.M. Davis, and C.H. Lineweaver, *Nature* **418**, 602-603 (2002)
[3] S.W. Hawking, *Phys.Rev*. **D13**, 191-197 (1976)
[4] S.W. Hawking, *Nature* **248**, 30-31 (1974)
[5] S.W. Hawking, *Commun.Math.Phys.* **43**, 199-200 (1975)
[6] D.N. Page, *Phys.Rev*. **D13**, 198-206 (1976)
[7] D.N. Page, *Phys.Rev*. **D14**, 3260-3273 (1976)
[8] J.D Bekenstein, *Phys.Rev*. **D12**, 3077-3085 (1975)
[9] J.D. Bekenstein, in *Proceedings of the Seventh Marcel Grossman Meeting on General Relativity* (World Scientific, Singapore, 1994), pp 39-58